\documentclass{aastex}          
\usepackage{spr-astr-addons}    

 \newcommand{\fgl}{2FGL~J2056.7+4939}

\begin{document}
%
\title{Sub-arcsecond radio and optical observations of the likely counterpart to the gamma-ray source \fgl}

\shorttitle{The radio counterpart to \fgl}
\shortauthors{Mart\'{\i} et al.}

\author{J. Mart\'{\i}}
\affil{Departamento de F\'{\i}sica, Escuela Polit\'ecnica Superior de Ja\'en, Universidad de Ja\'en, Campus Las Lagunillas A3-420, 23071 Ja\'en, Spain}
\and 
\author{P. L. Luque-Escamilla}
\affil{Departamento de Ingenier\'{\i}a Mec\'anica y Minera, Escuela Polit\'ecnica Superior de Ja\'en, Universidad de Ja\'en, Campus Las Lagunillas A3-008, 23071 Ja\'en, Spain}
\and
\author{E. S\'anchez-Ayaso}
\and
\author{A. J. Mu\~noz-Arjonilla}
\affil{Departamento de F\'{\i}sica, Escuela Polit\'ecnica Superior de Ja\'en, Universidad de Ja\'en, Campus Las Lagunillas A3-065, 23071 Ja\'en, Spain}


\begin{abstract}
We have searched and reviewed all multi-wavelength data available for the region towards the gamma-ray source \fgl\
in order to constrain its possible counterpart at lower energies. As a result, only a point-like optical/infrared source
with flat-spectrum radio emission is found
to be consistent with all X-ray and gamma-ray error circles. Its structure is marginally resolved at radio wavelengths
at the sub-arcsecond level.
An extragalactic scenario appears to be the most likely
interpretation for this object.
\end{abstract}


%
\section{Introduction}

The low galactic latitude gamma-ray source 
known as
 \fgl\ is one of the entries in the recent  Large Area Telescope (LAT) 2-year Point Source Catalog
provided by the Fermi Gamma-ray Space Telescope  in the 100 MeV to 100 GeV energy range \citep{fermi}. The Fermi
team points out a preliminary classification as an active galactic nucleus (AGN) of unknown type. This high-energy source is likely
to have been detected as well by other observatory missions in the past in both X-rays and soft gamma-rays. The reader
is referred to Table \ref{conioya} for an historical account. Attempts to find out the nature of this object based on these lower energy
detections have provided no conclusive result yet. In particular, inside the \fgl\ 95\% confidence ellipse there is only one conspicuous X-ray
source, namely 1RXS J205644.3+494011. Up to very recently, the identification of this ROSAT source with the luminous star BD+49 3420
was still considered plausible instead of an extragalactic origin \citep{hr09}. In addition, \citet{pmr} proposed this X-ray emitter as a possible microquasar 
candidate pointing out its coincidence with an intense radio source. 

In this work, we address the new Fermi detection together with all the multi-wavelength observational data available to try to shed light about the 
true origin of \fgl.


%
\begin{figure}[tb]
 \includegraphics[width=\columnwidth]{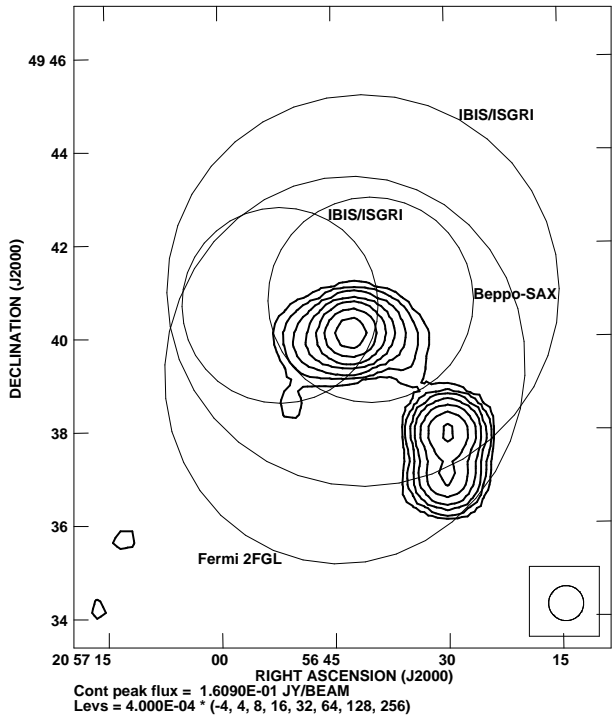}
 \caption{Radio sources from the NVSS in the direction of the \fgl\ field of view. The angular resolution is
 given by the beam size of $45^{\prime\prime}$ shown at the bottom right corner. The 90\%
 confidence error circles of sources detected by Fermi, INTEGRAL IBIS/ISGRI (soft-gamma and hard-X rays)
 and Beppo-SAX are also plotted. NVSS J205642+494005
 at the map centre is the only source consistent with all these error circles} 
 \label{fig1}
 \end{figure}

\section{Radio observations}

Radio imaging of high energy sources has proved to be an important tool to their understanding.
In Fig. \ref{fig1}, we show a contour map of radio emission in the field of \fgl\ according to the NRAO VLA Sky Survey (NVSS, \citet{nvss}).
Only the extended radio source  NVSS J205642+494005 is consistent even with the less accurate positions in Table \ref{conioya}
reported by Fermi, INTEGRAL and Beppo-Sax. As the NVSS has a poor angular resolution, we inspected the NRAO data archives
in search for interferometric observations with long baselines. We found only one observing run suitable for our purposes under project
code AK0360 and carried out at the 6 cm wavelength on 1994 May 8th, with the Very Large Array (VLA) in its hybrid AB configuration.
The data were retrieved and calibrated using the AIPS package of NRAO. The source 3C286 was used for amplitude calibration, while
the phase calibrator was the nearby source 2202+422. Finally, self-calibration was applied as the target source was bright enough.

\begin{figure}[tb]
 \includegraphics[width=\columnwidth]{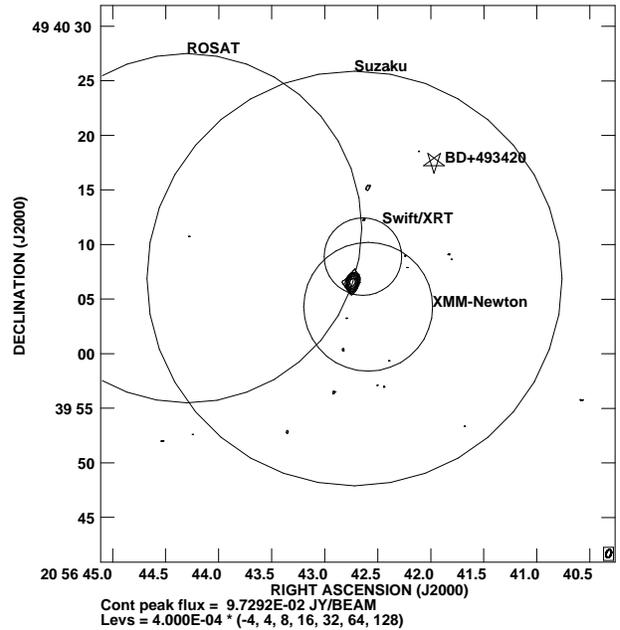}
 \caption{High angular resolution VLA map at the 6 cm wavelength showing the
 compact radio emission from the direction of \fgl\ and coincident with the core of NVSS J205642+494005 in Fig. \ref{fig1}.
The bottom right ellipse shows the synthesized beam and corresponds to $0.76 \times  0.45$ arc-second$^2$,
 with position angle of $-18$ deg. The error circles of ROSAT, Suzaku, Swift/XRT and XMM-Newton X-ray detections
 in the field are also shown. The star symbol corresponds to the location of the nearby, luminous star BD+49 3420} 
 \label{fig2}
 \end{figure}

As a result, we obtained the contour map shown in Fig. \ref{fig2} revealing that the extended NVSS source also exhibits
compact emission components.
This was achieved using a pure uniform weighting of the interferometer visibilities (AIPS ROBUST parameter $-5$), thus yielding a
sub-arcsecond angular resolution. NVSS J205642+494005 appears as partially resolved. An elliptical Gaussian fit using
the AIPS task JMFIT gives a deconvolved angular size of
$(290\pm 11) \times (173 \pm 14)$ mas$^2$, with a position angle of $17\pm5^{\circ}$. The radio source is centered at the J2000.0 position
$20^h 56^m 42.738^s$ and $+49^{\circ} 40^{\prime} 06.55^{\prime\prime}$, with an estimated uncertainty of about $0.01^{\prime\prime}$ 
in each coordinate. This is fully consistent with the most precise error circles of Table \ref{conioya}, as is also illustrated in  Fig. \ref{fig2}.
The field of view of this second figure zooms deeply on the central parts of Fig. \ref{fig1} and the most accurate error circles of
Table \ref{conioya} are overlayed on it.
The total integrated flux density of NVSS J205642+494005
 at 6 cm is $117 \pm 1$  mJy. Although we cannot accurately estimate the spectral index, consultation
of the SPECFIND v2.0 catalogue \citep{specfind} suggests that this object is a flat-spectrum radio source.

\begin{figure}[tb]
 \includegraphics[width=\columnwidth]{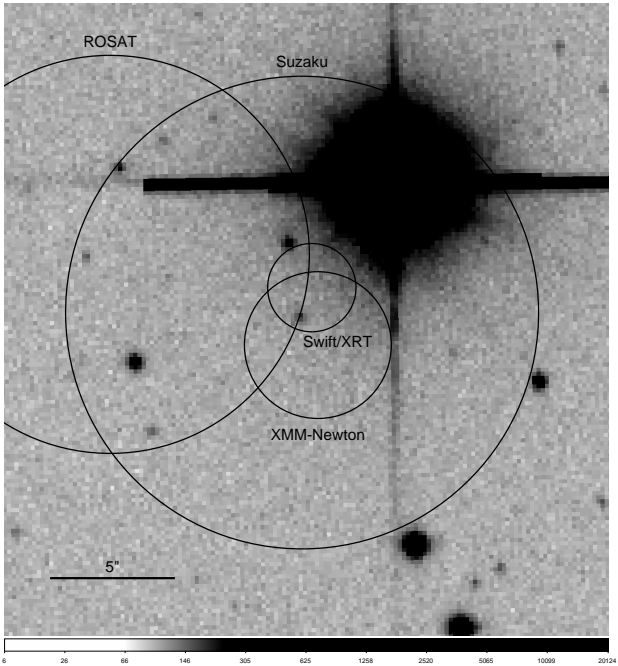}
 \caption{Same field of view of Fig. \ref{fig2} as it appears in IPHAS data taken in the optical $i'$-band under good seeing conditions. A single, clearly point-like 
 object is consistent with the VLA radio source. The nearby bright, saturated star is BD+49 3420. The horizontal bar
 indicates the angular scale. North is up and East is left} 
 \label{fig3}
 \end{figure}

\section{Optical and infrared data}

The infrared counterpart of NVSS J205642+494005 has been already pointed out by \citet{l10}
based on the 2 Micron All Sky Survey (2MASS),and is designated as  2MASS J20564271+4940068. It is located at a J2000.0 position
$20^h 56^m 42.72^s$ and  
$+49^{\circ} 40^{\prime} 06.9^{\prime\prime}$
and with magnitudes
$J = 13.7$, $H=14.3$ and $K=13.7$. When inspecting the original 2MASS images, one finds that
the detection of this infrared source is severely perturbed by the presence of the extremely bright nearby star BD+49 3420.

\begin{table*}[tb]
\tiny
\caption{Archival detections towards the direction of \fgl} 
\label{conioya}
\begin{tabular}{lllllll}
\tableline  
 
                    Instrument                                                     & 
                   Energy range                                                 & 
                   Source Id.                                                      & 
                   R.A. (J2000.0)                                               &  
                   DEC. (J2000.0)                                              & 
                   Position                                                          &  
                   Reference                                                      \\
                                                                                          & 
                                                                                          & 
                                                                                          & 
                   {h}\phn{m}\phn{s}                                           &  
                   \phn{\arcdeg}~\phn{\arcmin}~\phn{\arcsec}   & 
                   error                                                               \\ 
\hline                   
   ROSAT           & 0.1-2.4 keV  & 1RXS  J205644.3+494011        & 20 56 44.30   & +49 40 11.5  & $16^{\prime\prime}$    &  \cite{v99} \\
   Swift/XRT       & 0.3-10 keV   &  SWIFTJ2056.8+4939               &  20 56 42.64  & +49 40 08.9  & $3.55^{\prime\prime}$ &  \cite{l10} \\
   XMM-Newton & 0.2-12 keV   &  XMMSL1 J205642.7+494004   & 20 56 42.59  & +49 40 04.3  & $5.9^{\prime\prime}$\tablenotemark{a}  & \cite{xmmslew} \\ 
   Beppo-SAX    & 3-17 keV      & IGR J20569+4940                     & 20 56 40     & +49 40.8         & $2.2^{\prime}$               & \cite{sax} \\
   IBIS/ISGRI     & 17-100 keV  &  IGR J20569+4940                    &  20 56 41   & +49 41.0          & $4.2^{\prime}$              & \cite{4th-ibis-isgri} \\
                          & 17-60 keV    &                                                   &  20 56 52   &  +49 40.7        &  $2.1^{\prime}$              & \cite{ibis} \\
   Suzaku           &  0.2-600 keV  & RXJ2056.6+4940                    & 20 56 42.72 & +49 40 06.9  & $19^{\prime\prime}$           & \cite{suzaku}\\
    Fermi             & 100 MeV-100 GeV &   2FGL J2056.7+4939     &  20 56 43.5 &  +49 39.3  & $4^{\prime}$ &   \cite{fermi} \\        

\tableline 
\end{tabular}
%
 \tablenotetext{a}{\tiny 1 $\sigma$ error.}
\end{table*}

To search for an optical counterpart, so far unreported, we used one of the newest and deepest optical
surveys of the region available nowadays,
such as the INT Photometric H$\alpha$ Survey (IPHAS) of the Northern Galactic Plane in its $r^{\prime}$ and
$i^{\prime}$ broad-band filters \citep{iphas}. The IPHAS images covering the \fgl\ position were taken under very good
sub-arcsecond seeing conditions. Fig. \ref{fig3} shows the same field as the high angular resolution radio map of Fig.\ref{fig2}
as observed in the $i^{\prime}$ filter with a seeing value of $0.6^{\prime\prime}$. Here, a clearly stellar-like optical source is also
consistent with the radio core of NVSS J205642+494005  and the 2MASS object within a few tenths of arc-second. The magnitudes of
the IPHAS object, which is designated as IPHAS J205642.74+494006.7, are estimated as $i^{\prime} = 19.5$ and $r^{\prime} \leq 21$.

\section{Discussion and conclusions}

The coincidence in position within astrometric error of the X-ray source 1RXS J205644.3+494011, the flat-spectrum radio emitter
NVSS J205642+494005, the point-like infrared and optical sources
2MASS J20564271+4940068 / IPHAS J205642.74+494006.7
and all high energy error boxes listed in Table \ref{conioya}, strongly points to all being the same
astronomical source.
This is in addition the most peculiar object within the error ellipse of \fgl.
The possible connection with this gamma-ray source is reinforced after running a Monte Carlo simulation (see the methodology in \cite{romero99})
to estimate the probability of a chance coincidence between the X-ray and the Fermi sources, that turns out to be of the order of 0.4\%. 
On the other hand, the bright star BD+49 3420 is clearly excluded as candidate counterpart due to its non positional coincidence
with most X-ray detections.

The data currently available are not able to clearly find out the nature, galactic or extragalactic, of the proposed counterpart to \fgl.
Nevertheless, a crude assessment is possible relying on broad band photometry.
Based on X-ray spectral fits, the total hydrogen column density towards it
is estimated as about $5.3 \times 10^{21}$ cm$^{-2}$ \citep{l10}. This number translates into a visual extinction of
$A_V \simeq 3.9$ mag or, equivalently, a colour excess of $E(B-V) \simeq 1.3$. Taking this into account, the deredened 
photometric colors of the IPHAS/2MASS counterpart amount to:
$r^{\prime} - i^{\prime} \geq 0.7$, $J-H\simeq -0.9$, $J-K \simeq -0.6$, and $H-K\simeq 0.4$. 
When checking against tables of intrinsic stellar colours (e.g. \citet{iphas,rm91}), we find that the IPHAS colour lower limit
would be consistent with a main sequence star later than A2V. However, the infrared colours cannot be easily matched with
any kind of normal star. Therefore, the spectrum of this source does not resemble to be stellar and the possibility
of being an extragalactic object is thus reinforced. If this is the case, the point-like appearance of the optical/infrared counterpart
would be typical of a distant AGN (possibly a quasar or a blazar). In this case, the observed flat radio spectrum could be
naturally interpreted in terms of self-absorption of synchrotron radio photons in the compact core.

There exists another outstanding radio source inside the error circle of \fgl\ as may be seen in Fig. \ref{fig1}. 
Its extended appearance at 20 cm wavelength suggests a possible one-sided jet structure. 
The 1994 VLA archival data used in this paper clearly detect the position
of a compact core consistent with NVSS J205629.93+493756.4. Its precise J2000.0 coordinates are 
$20^h 56^m 30.25^s$ and $+49^{\circ} 37^{\prime} 59.3^{\prime\prime}$, with an estimated uncertainty of about $0.1^{\prime\prime}$
with a measured  6 cm flux density of  $21 \pm 3$ mJy. In addition, the spectral index seems to
be non-thermal according to  \citet{specfind}. 
However, despite all these facts the identification of this NVSS source with \fgl\ seems much less likely
than for the NVSS J205642+494005. The key reason is the absence of any hints of X-ray emission which one would normally
expect as a by-product in the context of common physical scenarios with production of gamma-rays.

To conclude,  we have reported new radio and optical observations at sub-arcsecond scales towards the recently detected Fermi source \fgl. 
Together with archival data at X-ray wavelengths, we propose that the 2MASS J20564271+4940068/IPHAS J205642.74+494006.7 
is the most plausible candidate counterpart of this gamma-ray source
instead of the luminous star BD+49 3420, as sometimes still quoted in the literature. 
Further optical/infrared spectroscopic observations would be required to find the true nature of this object, although current evidence
suggests that an extragalactic scenario is a likely one.



%
%

%
%

%


%

%
\acknowledgments
The authors acknowledge support by grants AYA2010-21782-C03-03 from the Spanish Government, Consejer\'ia de Econom\'ia, Innovaci\'on y Ciencia of Junta de Andaluc\'ia as research group FQM-322 and excellence fund FQM-5418, as well as FEDER funds. The NRAO is a facility of the NSF operated under cooperative agreement by Associated Universities, Inc. This research made use of the SIMBAD database, operated at the CDS, Strasbourg, France. This publication makes use of data products from the Two Micron All Sky Survey, which is a joint project of the University of Massachusetts and the Infrared Processing and Analysis Center/California Institute of Technology, funded by the National Aeronautics and Space Administration and the National Science Foundation in the USA. 


%

%

\end{document}